\documentclass[pdflatex,sn-mathphys-num]{sn-jnl}% Math and Physical Sciences Numbered Reference Styl

\usepackage{graphicx}%
\usepackage{multirow}%
\usepackage{amsmath,amssymb,amsfonts}%
\usepackage{amsthm}%
\usepackage{mathrsfs}%
\usepackage[title]{appendix}%
\usepackage{xcolor}%
\usepackage{textcomp}%
\usepackage{manyfoot}%
\usepackage{booktabs}%
\usepackage{algorithm}%
\usepackage{algorithmicx}%
\usepackage{algpseudocode}%
\usepackage{listings}%
\usepackage{placeins}
\usepackage{tabularx}
\usepackage[mathlines]{lineno}

\theoremstyle{thmstyleone}%
\theoremstyle{thmstyletwo}%

\theoremstyle{thmstylethree}%

\begin{document}

%\linenumbers

\title[Article Title]{Quantitative mapping of dynamic 3D transport in growing cells via volumetric spatio-temporal image
correlation spectroscopy (vSTICS)}

\author[1]{\fnm{Ahmad} \sur{Mahmood}}

\author[1,2]{\fnm{Paul W.} \sur{Wiseman}}\email{paul.wiseman@mcgill.ca}

\affil[1]{\orgdiv{Department of Physics}, \orgname{McGill University}, \orgaddress{\street{3600 Rue University}, \city{Montreal}, \postcode{H3A 2T8}, \state{Quebec}, \country{Canada}}}

\affil[2]{\orgdiv{Department of Chemistry}, \orgname{McGill University}, \orgaddress{\street{801 Sherbrooke Street West}, \city{Montreal}, \postcode{H3A 0B8}, \state{Quebec}, \country{Canada}}}

% *** EDITED: structured, methods-first abstract
\abstract{\textbf{Purpose:} Quantitatively mapping three-dimensional (3D) flow, diffusion, 
and particle density in crowded living cells remains challenging: most dynamic 
optical microscopy measurements are effectively planar and existing analysis 
methods struggle with dense, noisy volumetric data. We introduce volumetric 
spatio-temporal image correlation spectroscopy (vSTICS), a general framework 
that recovers voxel-resolved flow, diffusion coefficients, and particle densities 
from 3D fluorescence time series. We validate vSTICS using simulated flow--diffusion 
data and fluorescent microsphere calibrations, then apply it to mitochondria and 
sub-diffraction vesicles in growing \emph{Camellia japonica} pollen tubes as a 
stringent biological test case.

\textbf{Methods:} We rapidly acquire 3D time series of growing \emph{Camellia japonica} 
pollen tubes using field-synthesis lattice light-sheet microscopy to maximize imaging 
speed and minimize photobleaching. A localized 3D spatio-temporal correlation analysis 
is applied to overlapping volumetric samples, yielding voxel-resolved velocity vectors, 
diffusion coefficients, and density estimates. Validation uses synthetic flow--diffusion 
simulations and fluorescent microsphere samples spanning practical imaging conditions.

\textbf{Results:} Simulations demonstrated accurate recovery of seeded transport parameters 
(velocities near 3~\textmu m\,s\textsuperscript{$-1$} and diffusion around 
$10^{-3}$~\textmu m\textsuperscript{2}\,s\textsuperscript{$-1$}). Microsphere experiments 
verified particle number and point spread function readouts and measured diffusion coefficients 
of $0.3\pm0.1$~\textmu m\textsuperscript{2}\,s\textsuperscript{$-1$} in gel, consistent with 
imaging-FCS measurements ($0.5\pm0.2$~\textmu m\textsuperscript{2}\,s\textsuperscript{$-1$}). 
For mitochondrial dynamics in pollen tubes, vSTICS resolved a bidirectional ``reverse-fountain'' 
with slower anterograde transport (0.1--1~\textmu m\,s\textsuperscript{$-1$}) and faster retrograde 
motion peaking near 3~\textmu m\,s\textsuperscript{$-1$}. Transverse flow profiles exhibited 
a characteristic retrograde corridor width of about 2~\textmu m. Density and diffusion maps 
revealed a denser, more advective core (6 vs.\ 4~particles\,\textmu m\textsuperscript{$-3$}; 
center vs.\ periphery) and systematically higher peripheral diffusion coefficients 
(e.g.\ median values elevated relative to the core). High-density, sub-diffraction-sized 
vesicle mapping yielded similar velocity landscapes to the mitochondria, but with roughly 
ten-fold higher particle densities.

\textbf{Conclusion:} vSTICS enables quantitative 3D mapping of intracellular transport with 
practical parameterization guidelines for volume size, overlap, and time-lag sampling. The method 
refines the reverse-fountain model by revealing asymmetric, predominantly transverse circulation, 
and provides transferable workflows for quantitative volumetric light-sheet fluorescence imaging 
studies beyond pollen tubes.}


\maketitle

\section{Introduction}\label{Introduction}

Three-dimensional (3D) optical sectioning fluorescence microscopy has emerged as a key methodology in biological research, providing an important way to image the dynamics of complex biological processes across a range of temporal and spatial scales. Techniques such as confocal microscopy \cite{Conchello2005}, multiphoton microscopy \cite{Denk1990}, and light-sheet fluorescence microscopy (LSFM) \cite{Huisken2004} have enabled the visualization of biological dynamics within intact living organisms, tissues, and cells. The advent of lattice light-sheet microscopy (LLSM) \cite{Chen2014} further advanced high spatio-temporal imaging capabilities by significantly reducing phototoxicity and photobleaching compared to conventional methods. Field-synthesis lattice light-sheet microscopy, a recent innovation, specifically minimizes photobleaching by shaping illumination fields dynamically, thus optimizing sample viability during extended volumetric imaging sessions \cite{Chang2019}, while also drastically simplifying the optical train. However, volumetric visualization becomes more powerful when paired with quantitative analysis of the dynamic structures.

A fundamental biological process captured effectively by 3D fluorescence microscopy imaging is molecular transport, primarily as active flow and passive diffusion. These transport mechanisms play critical roles in a diverse array of biological processes, ranging from intracellular vesicle trafficking and diffusion of biomolecules within the cytoplasm, to extracellular fluid or cell movements within tissues, and even larger-scale processes like blood flow or lymphatic drainage in vertebrates \cite{Swartz2007, Wiig2012}. For example, intracellular diffusion and directed transport via cytoskeletal elements determine critical signaling pathways and metabolic homeostasis in animal cells \cite{Vale2003, Fletcher2010}. In plants, targeted vesicle transport underlies rapid growth responses and directional elongation, as seen notably in pollen tube growth, essential for successful fertilization and reproduction \cite{Chebli2013}. Quantifying such transport phenomena accurately and robustly remains challenging due to limitations inherent in current analytical approaches. Single particle tracking (SPT), while powerful, is limited by difficulties in tracking individual particles in dense and dynamic environments, where high particle density and overlapping fluorescence signals often cause ambiguities in trajectory reconstruction and particle identity fidelity \cite{Manzo2015}. Particle image velocimetry (PIV), another widely used method, performs well in estimating ensemble velocity fields but suffers from low sensitivity to diffusive processes and struggles with accurate velocity measurements in low signal-to-noise conditions and highly heterogeneous environments \cite{Estruch-Samper_2012}.

Spatio-temporal image correlation spectroscopy (STICS) has addressed some of these limitations by leveraging correlation analysis to quantify average transport dynamics from fluorescence intensity fluctuations, enabling efficient measurement of flow and diffusion from input fluorescence microscopy image time series. However, the original STICS has only been applied to measure transport in quasi two-dimensional (2D) systems, including protein velocity mapping in living cells, adhesion dynamics in migrating cells, and vesicular transport in plant cells, so most applications remain effectively planar despite the inherently volumetric nature of many biological structures \cite{Hebert2005, Toplak2012, Bove2008VesicleDynamicsPollenTubes}. 

Conventional techniques frequently measure velocities confined to a single imaging plane, even when applied to inherently volumetric biological processes. For instance, cytoplasmic streaming velocities in large plant cells have been reported in the range of 2$-$10 $\mu$m/s using 2D PIV and STICS techniques \cite{Shimmen2004, Woodhouse2013}. Similarly, blood flow velocities measured in microvascular networks using planar imaging approaches typically range from tens to hundreds of $\mu$m/s \cite{Jain2005}. Although these measurements offer valuable insights, their confinement to a single focal plane in inherently three-dimensional biological systems restricts comprehensive characterization of the underlying flow fields, potentially overlooking critical dynamics occurring outside the imaged plane and leading to incomplete or biased interpretations \cite{Ohnuki2021}.

Pollen tube elongation exemplifies a system where precise quantification of flow and diffusion dynamics is crucial. The rapid and directional elongation of pollen tubes involves coordinated transport of vesicles, organelles, and cytoplasmic components, guided by dynamic cytoskeletal networks and ion concentration gradients \cite{Rounds2011}. Efficient vesicle transport and delivery at the pollen tube tip are crucial to maintaining rapid growth rates, directly impacting reproductive success \cite{Hepler2001}. Precise, quantitative measurements of three-dimensional flow and diffusion in these tubes are therefore essential to validate theoretical and computational models of cell growth and morphology and to understand the mechanisms that govern plant reproduction. % *** EDIT: removed vSTICS here and ended with purely biological motivation

\begin{figure*}[t]
  \centering
  \includegraphics[width=0.825\textwidth]{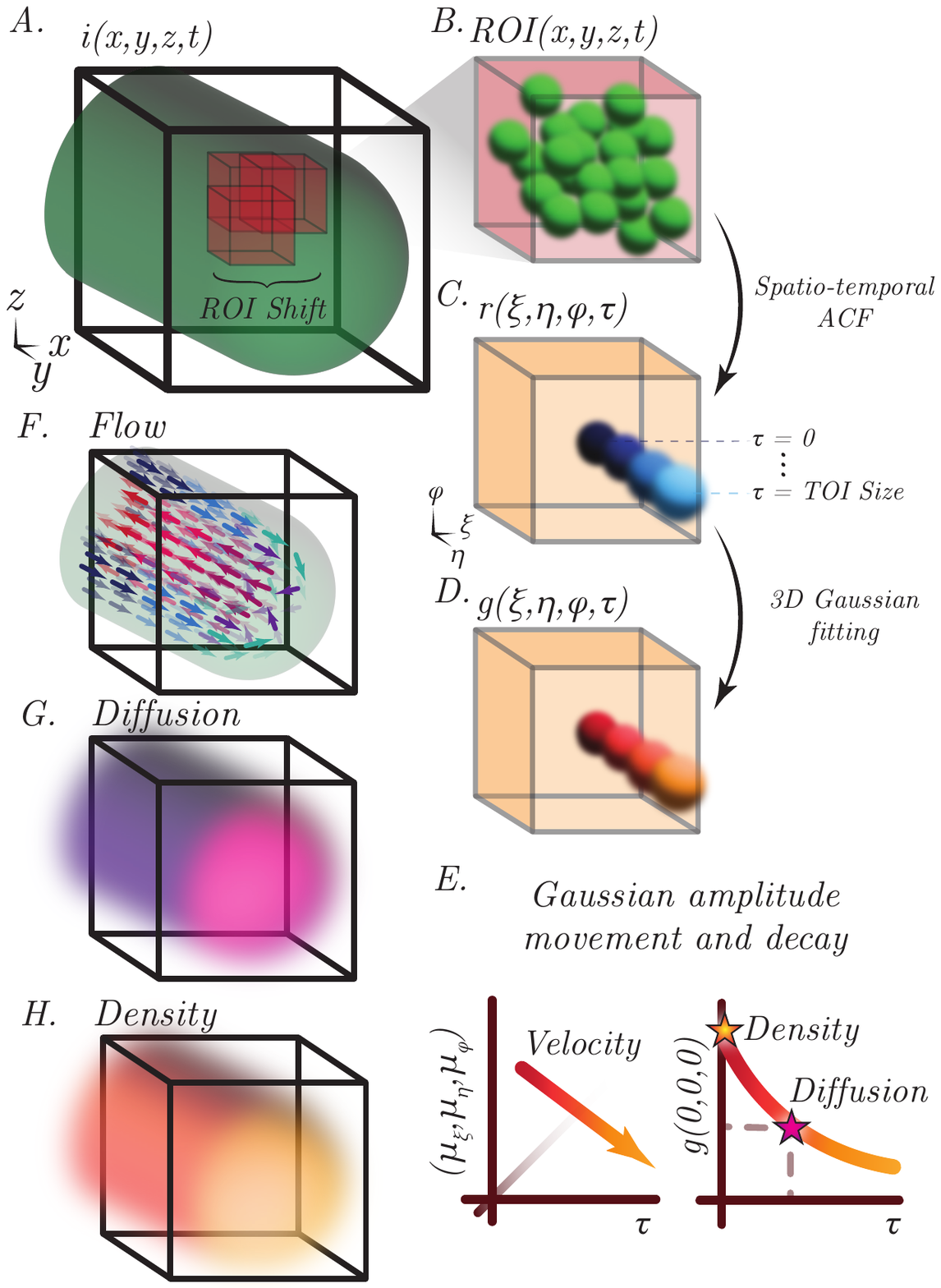}
  \caption{\textbf{Overview of volumetric STICS (vSTICS).}
  \textbf{(A)} A 4D image sequence, $i(x,y,z,t)$, is sampled with overlapping regions of interest (ROIs) that are shifted through the volume.
  \textbf{(B)} For each local ROI$(x,y,z,t)$, fluorescence fluctuations are analyzed over a finite time window.
  \textbf{(C)} The spatio-temporal autocorrelation function, $r(\xi,\eta,\phi,\tau)$, is computed for a sequence of lag times $\tau$, producing a correlation peak that evolves with transport.
  \textbf{(D)} At each lag, the correlation function is fit with an asymmetric 3D Gaussian model, $g(\xi,\eta,\phi,\tau)$.
  \textbf{(E)} The fitted Gaussian parameters summarize transport: the peak position $(\mu_{\xi},\mu_{\eta},\mu_{\phi})$ shifts with $\tau$ and yields the local velocity, while the fitted amplitude $g(0,0,0,\tau)$ reports density and diffusion. In particular, the zero-lag amplitude is inversely proportional to the average number of particles in the sampling volume, and its decay with $\tau$ reflects diffusive motion.
  \textbf{(F--H)} Repeating this analysis over all overlapping ROIs generates volumetric maps of \textbf{flow} (F), \textbf{diffusion} (G), and \textbf{density} (H).}
  \label{fig:overview}
\end{figure*}

We therefore developed volumetric spatio-temporal image correlation spectroscopy (vSTICS), a volumetric correlation workflow that estimates voxel-resolved velocity, diffusion, and density from 3D fluorescence time series (Fig.~\ref{fig:overview}). vSTICS takes volumetric fluorescence movies as input and, through localized 3D correlation and asymmetric Gaussian fitting, returns quantitative maps of flow, diffusion and density at user-defined spatial resolution, together with practical guidelines for choosing block sizes, overlaps and time-lag sampling across microscopes and specimens. We first validate vSTICS using simulations and fluorescent microsphere calibrations and then use it to quantify mitochondria and sub-diffraction vesicle transport in growing \emph{Camellia japonica} pollen tubes imaged with field-synthesis lattice light-sheet microscopy. Although we focus on pollen tubes here, the same framework is directly applicable to 3D transport in axons, cilia, glandular ducts, blood flow, and other geometrically constrained environments where advection and diffusion are tightly coupled.

\section{Results: validation and application of vSTICS}\label{ResultsMain}

\subsection{Simulations}\label{ResultsSimulations}

\begin{figure}[t]
  \centering
  \includegraphics[width=\textwidth]{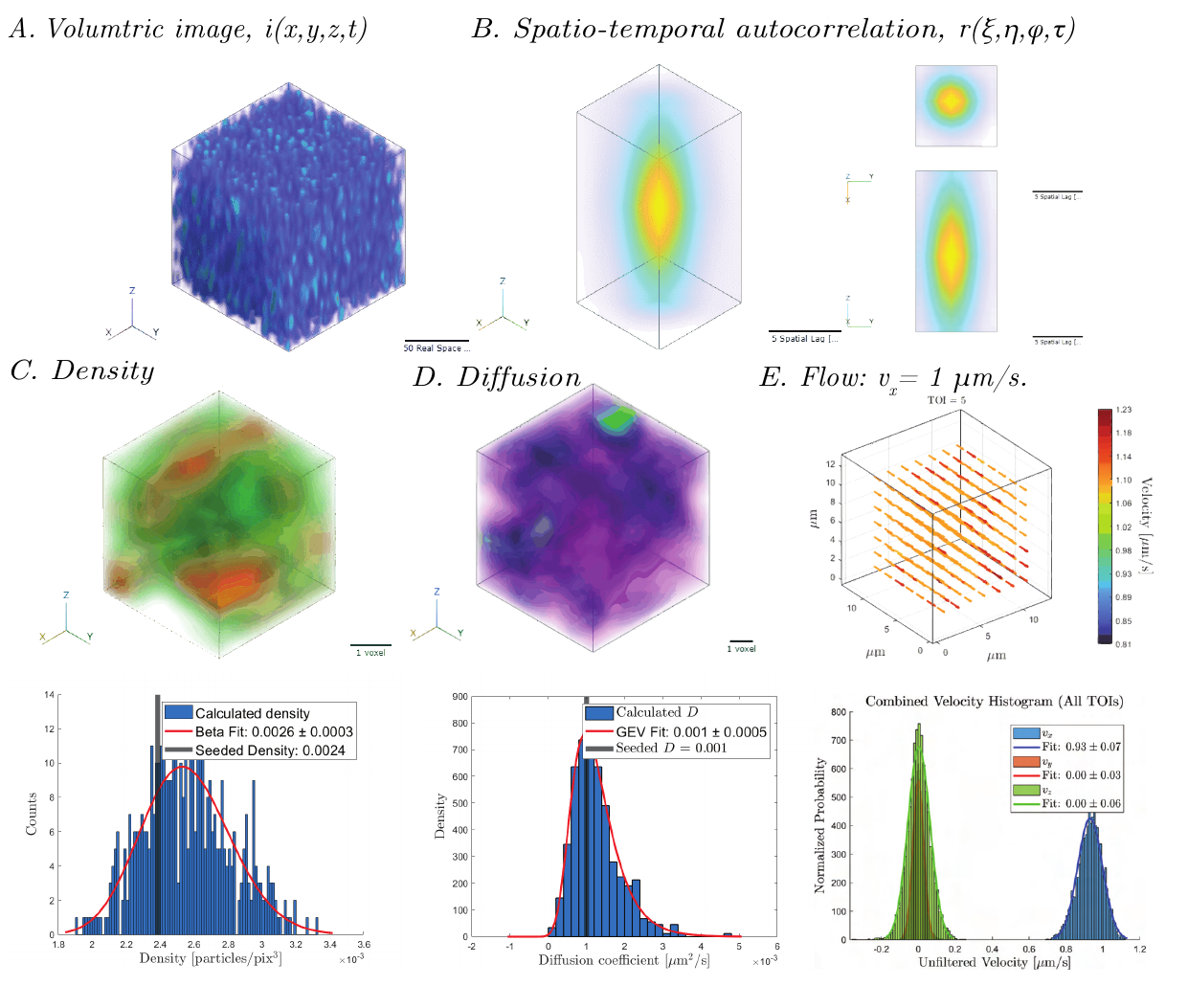}
  \caption{\textbf{Simulation benchmark for vSTICS.}
  \textbf{(A)} A volumetric image time series is simulated with an asymmetric Gaussian PSF with $\omega_{xy} = 1.5 $ pix and $\omega_{z} = 5$ pix. The image size is $128^{3} \ \text{pix}^{3}$ with a pixel size of 0.1 $\mu\text{m}$ for 128 time points, with an imaging time of 10 ms per volume.
  \textbf{(B)} The 3D normalized intensity fluctuation autocorrelation function at $\tau=0$, shown in a 3D view, as well as XY and XZ projections. A symmetric ROI size of 32 pix, with a ROI shift of 4 pix. 
  \textbf{(C)} The 3D density map of the volumetric time series at $\tau=1$ (\textit{top}) with its corresponding density histogram (\textit{bottom}) indicating both mean fitted value and seeded density.
  \textbf{(D)} The 3D diffusion map of the volumetric time series at $\tau=1$ (\textit{top}) with its corresponding diffusion histogram (\textit{bottom}) indicating both mean fitted value and seeded diffusion coefficient. A TOI window of 5 frames was used.
  \textbf{(E)} The 3D vector flow field of the volumetric time series at $\tau=1$ (\textit{top}) with its corresponding velocity histogram (\textit{bottom}) indicating both mean fitted value and seeded velocity in each Cartesian direction.}
  \label{fig:simulations}
\end{figure}

Using the custom simulation script (Section~\ref{Methods}: Computer simulations) to generate volumetric intensity time series (Fig.~\ref{fig:simulations}\textbf{A}), we validated vSTICS under three conditions modelling a single fluorescent species undergoing (i) pure flow, (ii) combined diffusion and flow, and (iii) combined diffusion and flow with additive white noise. The case of diffusion and flow is shown in Fig. \ref{fig:simulations} with velocity set to  $v_{x} = v_{y} = v_{z} = 3 \ \mu\text{m}/s$ and $D = 0.001 \ \mu\text{m}^{2}/s$, with a density of $2.4\cdot10^{-3} \ \text{particles}/\mu\text{m}^{3}$. The imaging parameters were set with an asymmetric Gaussian PSF with $\omega_{xy} = 1.5 $ pix and $\omega_{z} = 5$ pix. The image size was $128^{3} \ \text{pix}^{3}$ with a pixel size of 0.1 $\mu\text{m}$ and 128 time points, with an imaging time of 10 ms per volume. For volumetric correlation, a symmetric ROI size in $xyz$ of 32 pix, with a ROI shift of 4 pix, and a TOI window of 5 frames, with a shift of 1 frame was chosen  (Fig. \ref{fig:overview}: \textbf{A,B}). For each ROI, this sampling yields $5+1$ volumetric correlation functions (the additional term corresponding to the zero-lag correlation) to be fit with an asymmetric Gaussian function (Eq.~\ref{fitfunction}, Fig.~\ref{fig:overview}\textbf{D}). From the fitted correlation function, the peak position ($\mu_{\xi,\eta,\phi}$, Fig. \ref{fig:overview}: \textbf{E}) and amplitude ($g(\vec{0})$, Fig. \ref{fig:overview}: \textbf{F}) are recorded as a function of time lag $\tau$. The decay of the normalized correlation function amplitude provides a quantitative measurement of the diffusion coefficent, $D$, via the characteristic time decay which is the decorrelation time of the fluorescent molecule from the PSF defined observation volume (Fig. \ref{fig:overview}: \textbf{C}). By applying appropriate fit models to the decay \cite{Hebert2005} (Table S2, the density (Fig. \ref{fig:simulations}: \textbf{C}) and diffusion coefficient (Fig. \ref{fig:simulations}: \textbf{D}) can be extracted per TOI window for each ROI. By linear fitting  of the peak positions as a function of time lag (Fig. \ref{fig:overview}: \textbf{F}), velocity vector fields are generated per TOI window for each ROI (Fig. \ref{fig:simulations}: \textbf{E}). Using vSTICS, we recovered a density of $0.0028 \pm 0.0003~\text{particles}\,\mu\text{m}^{-3}$, 
a diffusion coefficient of $D = 0.0010 \pm 0.0005~\mu\text{m}^{2}\,\text{s}^{-1}$, 
and flow components of $v_x = v_y = 2.6 \pm 0.5~\mu\text{m}\,\text{s}^{-1}$ and 
$v_{z} = 2.1 \pm 0.8~\mu\text{m}\,\text{s}^{-1}$ under the sampling and simulation conditions reported.

\FloatBarrier

\subsection{Fluorescent microspheres}\label{ResultsFlMicro}

\subsubsection*{Volumetric spatial image correlation accurately measures particle numbers and PSF}

To validate the measurement accuracy of particle number with vSTICS, agar immobilized fluorospheres were imaged (Fig. \ref{fig:fluorospheres} \textbf{A}: \textit{left}). By calculating the normalized intensity fluctuation autocorrelation function (Fig. \ref{fig:fluorospheres} \textbf{A}: \textit{middle}), and fitting to  Eq. \ref{fitfunction} (Fig. \ref{fig:fluorospheres} \textbf{A}: \textit{right}), the amplitude and width are related to the number of particles and PSF respectively, as described in Section \ref{Theory}. The particle count was measured to be $N = 75$, and with image segmentation, $N = 63$, where image segmentation only measures full microspheres, where vICS will report on partial microspheres within the ROI crop as well. The structure factor, $\kappa = \omega_{z}/\omega_{xy}$, reported by vICS was $\kappa_{vICS} = 1.3 \pm 0.1$, and with segmentation, $\kappa_{seg} = 1.4 \pm 0.1$. This demonstrated both correct correlation function calculation, and near isotropic resolution achieved by the custom lattice light-sheet microscope.

\subsubsection*{Volumetric temporal image correlation accurately measures diffusion}

\begin{figure}[t]
  \centering
  \includegraphics[width=\textwidth]{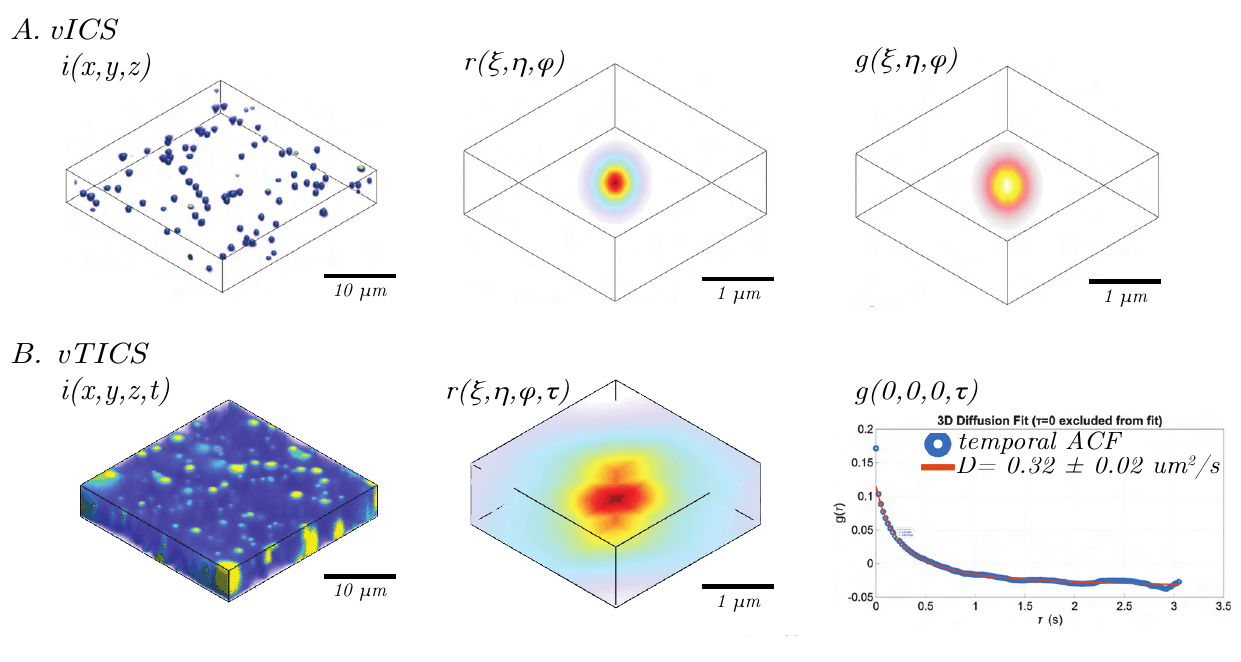}
\caption{\textbf{Fluorosphere calibration samples for experimental vSTICS validation.}
  \textbf{(A)} A 3D intensity stack of stationary fluorospheres of radius 50 nm, embedded in agarose with a field of view of 55 $\times$ 55 $\mu\text{m}^{2}$ (\textit{left}) with its corresponding volumetric normalized autocorrelation function (\textit{middle}) and its symmetric Gaussian fit (\textit{right}).
  \textbf{(B)} A 3D time series intensity stack of fluoropsheres of radius 50 nm in 0.3 $\%$ Gelrite mixture, with a field of view of 55 $\times$ 55 $\mu\text{m}^{2}$ (\textit{left}) with its corresponding volumetric normalized autocorrelation function at $\tau = 1$ (\textit{middle}) and the corresponding decay of the correlation amplitude $g(\vec{0)}$ as a function of time lag $\tau$, fit with a single component 3D diffusion model (\textit{right}).
  }
  \label{fig:fluorospheres}
\end{figure}

To validate transport measurements via vSTICS, volumetric correlation was performed, as previously shown in 2D \cite{Kolin2006TICS}. Sample bags of fluorescent microspheres were prepared as described in Section \ref{Methods}: Fluorosphere preparations. Fluorospheres of radius 50 nm have a theoretical Stokes-Einstein diffusion coefficient of $5 \ \mu\text{m}^{2}/s$ in water at 21 $^{\circ}$C. After increasing the dynamic viscosity coefficient of the solvent from 1 cP (water) to 8 cP (0.3$\%$ Gelrite mixture) \cite{Liu2010GelriteAlginate}, the theoretical diffusion coefficient decreases to $0.5 \ \mu\text{m}^{2}/s$. This enabled measurement of diffusion coefficients that lay within the volumetric imaging frequency bandwidth of the home built lattice light-sheet (Section \ref{Methods}: Lattice light-sheet microscopy via field synthesis). A volumetric time series measurement was taken (Fig. \ref{fig:fluorospheres} \textbf{B}: \textit{left}), and the temporal evolution of the volumetric spatial autocorrelation function (Fig. \ref{fig:fluorospheres} \textbf{B}: \textit{middle}) behaves as expected, with the time dependent expansion of the width. By correlating the time series strictly along the time dimension, the decay of the correlation peak can be fit with a one component diffusion model (Table S2), to yield a measured diffusion coefficient of $D_{vTICS} = 0.3 \pm 0.1 \ \mu\text{m}^{2}/s$. This measurement was compared to imaging fluorescence correlation spectroscopy (imFCS) measurements \cite{Bag2014IFFS} (Fig. S9), which yielded $D_{imFCS} = 0.5\pm 0.2 \ \mu\text{m}^{2}/s$ .

\FloatBarrier

\subsection{Dynamic volumetric mapping of mitochondria transport properties in \textit{in vitro} pollen tubes}\label{ResultsMito}

\begin{figure}[t]
  \centering
  \includegraphics[width=\textwidth]{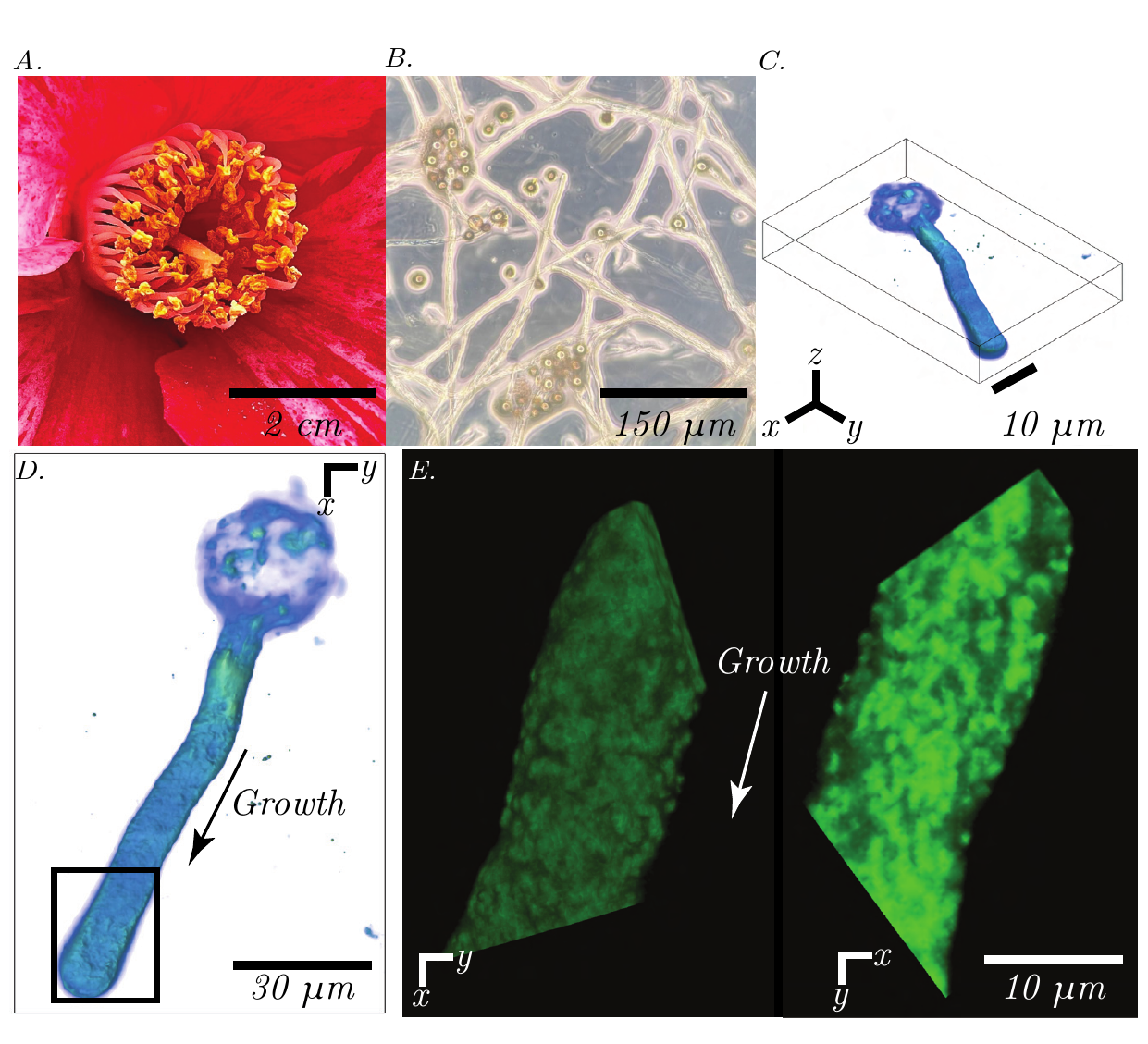}
  \caption{\textbf{\textit{In vitro} fertilization and mitochondrial labelling of \textit{Camellia japonica} pollen tubes.}
  \textbf{(A)} \textit{Camellia japonica} flower with pollen shown in yellow.
  \textbf{(B)} A bright field image of fertilized pollen grains grown on germination media (Section \ref{Methods}) 3 hours post deposition.
  \textbf{(C)} Lattice light-sheet image of a single pollen tube grown in a custom sample holder 2 hours post deposition, labeled with MitoTracker Green, and $xy$ projection shown in \textbf{(D)}. Region near the growing tip highlighted in the black box is shown in 
  \textbf{(E)}, with a bottom surface perspective of the tube (\textit{left}), and the sagittal plane of the tube (\textit{right}).}
  \label{fig:mitotrackergreen}
\end{figure}

\begin{figure}[t]
  \centering
  \includegraphics[width=\textwidth]{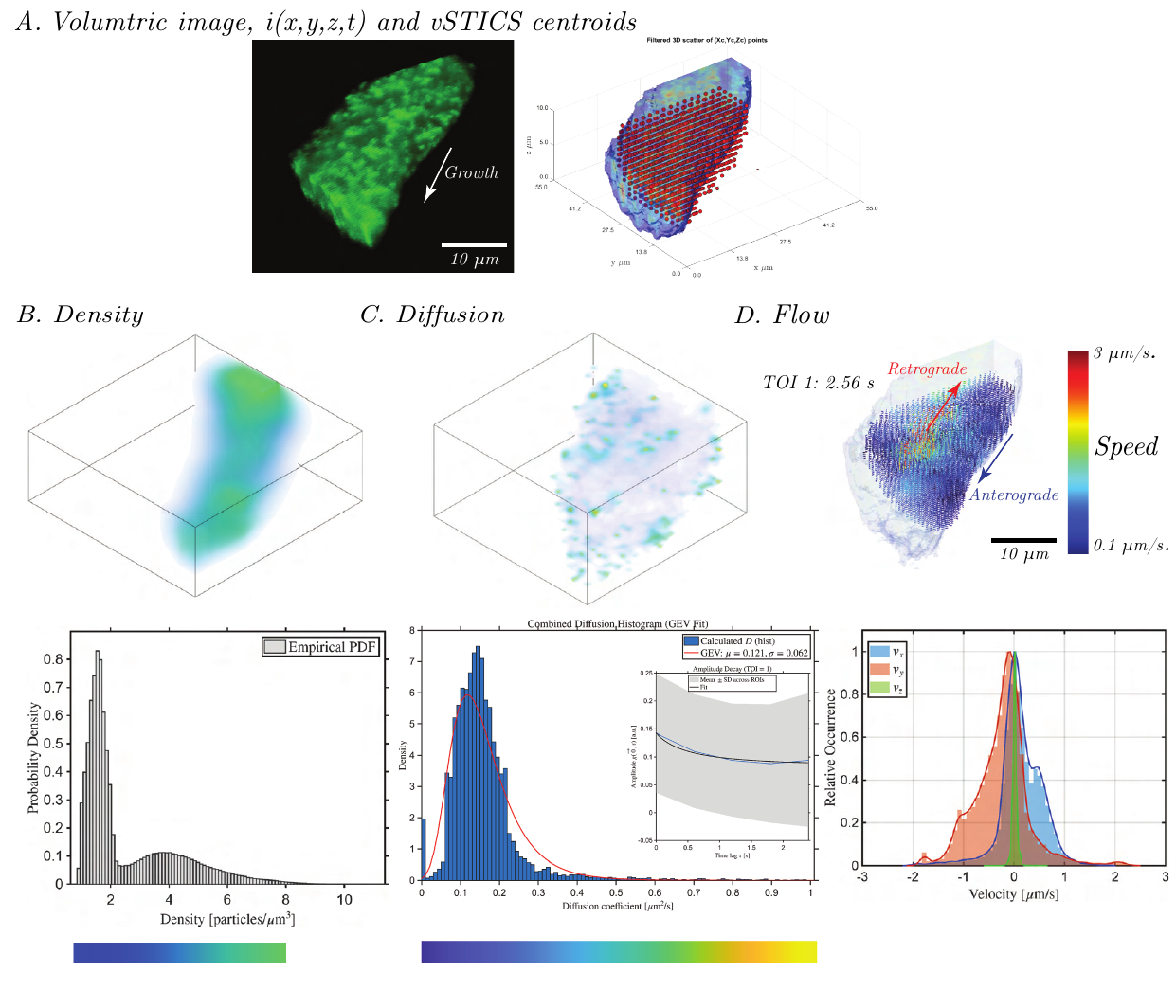}
  \caption{\textbf{Biological benchmark for vSTICS using mitochondrial labelling of \textit{Camellia japonica} pollen tubes.}
  \textbf{(A)} A volumetric time series of the pollen tube, displayed with the sagittal plane relative to the growth axis (\textit{left}), with discrete centroids of the $\text{ROIs}$ displayed in red, with parameters $\text{ROI}_{xy} = 64 \ \text{pix}$, $\text{ROI}_{z} = 16 \ \text{pix}$, $\text{ROI Shift}_{xy} = 5 \ \text{pix}$, $\text{ROI Shift}_{z} = 2 \ \text{pix}$, $\text{TOI} = 5 \ \text{frames}$, $\text{TOI Shift} = 1 \ \text{frame}$.
  \textbf{(B)} Volumetric density mapping of the growing pollen tube at the first $\text{TOI}$ (\textit{top}) with the density histogram averaged over all $\text{TOIs}$ with a Gaussian fit.  
  \textbf{(C)} Volumetric diffusion mapping of the growing pollen tube at the first $\text{TOI}$ (\textit{top}) with the diffusion histogram  (\textit{top}) averaged over all $\text{TOIs}$ with generalized extreme value (GEV) fit with an average $D = 0.15\ \mu\text{m}^{2}\,\text{s}^{-1}$. The inset show the decay of the ACF amplitude fit with a 3D diffusion model.
  \textbf{(D)} Velocity flow field at the first $\text{TOI}$ superimposed on a volumetric rendering of the pollen tube (\textit{top}). The corresponding normalized histogram of the velocity measured in each orthogonal direction, with a bimodal Gaussian fit for components $v_{x}$ and $v_{y}$, and a Gaussian fit for $v_{z}$.
  }
  \label{fig:vSTICSCJ}
\end{figure}

To test vSTICS on a complex, noisy biological volume and assess its robustness in the presence of biological variability, we applied it to mitochondrial dynamics within growing pollen tubes (Fig.~\ref{fig:mitotrackergreen} \textbf{A--C}). % *** EDIT: method-first framing
The established reverse-fountain mechanism for sperm delivery in pollen tubes provides a 3D living system in which vSTICS can interrogate the bidirectional flow (Fig.~\ref{fig:vSTICSCJ} \textbf{D--E}) at high spatio-temporal resolution (55~\textmu m ($200$~px) $\times$ 55~\textmu m ($200$~px) $\times$ 10~\textmu m ($100$~px); 
0.5~s per volume; 125 time points).
.

\FloatBarrier

Analysis was performed by oversampling the volumetric image time series  (Fig. \ref{fig:vSTICSCJ} \textbf{A}: \textit{left}) at discrete points in space (Fig. \ref{fig:vSTICSCJ} \textbf{A}: \textit{right}), with vSTICS analysis parameters set as: $\text{ROI}_{xy} = 64 \ \text{pix}$, $\text{ROI}_{z} = 16 \ \text{pix}$, $\text{ROI Shift}_{xy} = 5 \ \text{pix}$, $\text{ROI Shift}_{z} = 2 \ \text{pix}$, $\text{TOI} = 5 \ \text{frames}$, $\text{TOI Shift} = 1 \ \text{frame}$. Density mapping via vSTICS amplitude analysis (\ref{fig:vSTICSCJ} \textbf{B}: \textit{top}) provides the average spatial distribution per $\text{TOI}$ as the pollen tube grows with an average of $5 \pm 1 \ \text{particles/}\mu\text{m}^{3}$ (\ref{fig:vSTICSCJ} \textbf{B}: \textit{bottom}), with a higher concentrations ($6 \ \text{particles/}\mu\text{m}^{3}$) in the center, and lower concentrations at the periphery ($4 \ \text{particles/}\mu\text{m}^{3}$). Diffusion mapping provides an insight on the degree of Brownian motion undergone by particles in a spatially varying flow field (\ref{fig:vSTICSCJ} \textbf{C}: \textit{top}). The diffusion coefficients range from 0.1 to 1 $\mu\text{m}^{2}/s$, with higher diffusion coefficients at the periphery, and lower diffusion coefficients in the center (\ref{fig:vSTICSCJ} \textbf{C}: \textit{bottom}). Velocity vector field mapping (\ref{fig:vSTICSCJ} \textbf{D}: \textit{top}) highlights a key capability of vSTICS with accurate maps of the bidirectional flow for this dense complex system. Anterograde motion ranges from 0.1 to 1~\textmu m\,s\textsuperscript{$-1$}, while the retrograde flow is several times faster, with a peak near 3~\textmu m\,s\textsuperscript{$-1$} ($n_{\text{biological replicates}} = 5$). 

\FloatBarrier

\subsection{Dynamic volumetric mapping of sub-diffraction limited cargo vesicle transport properties in \textit{in vitro} pollen tubes}\label{ResultsVesicle}

\begin{figure}[t]
  \centering
  \includegraphics[width=\textwidth]{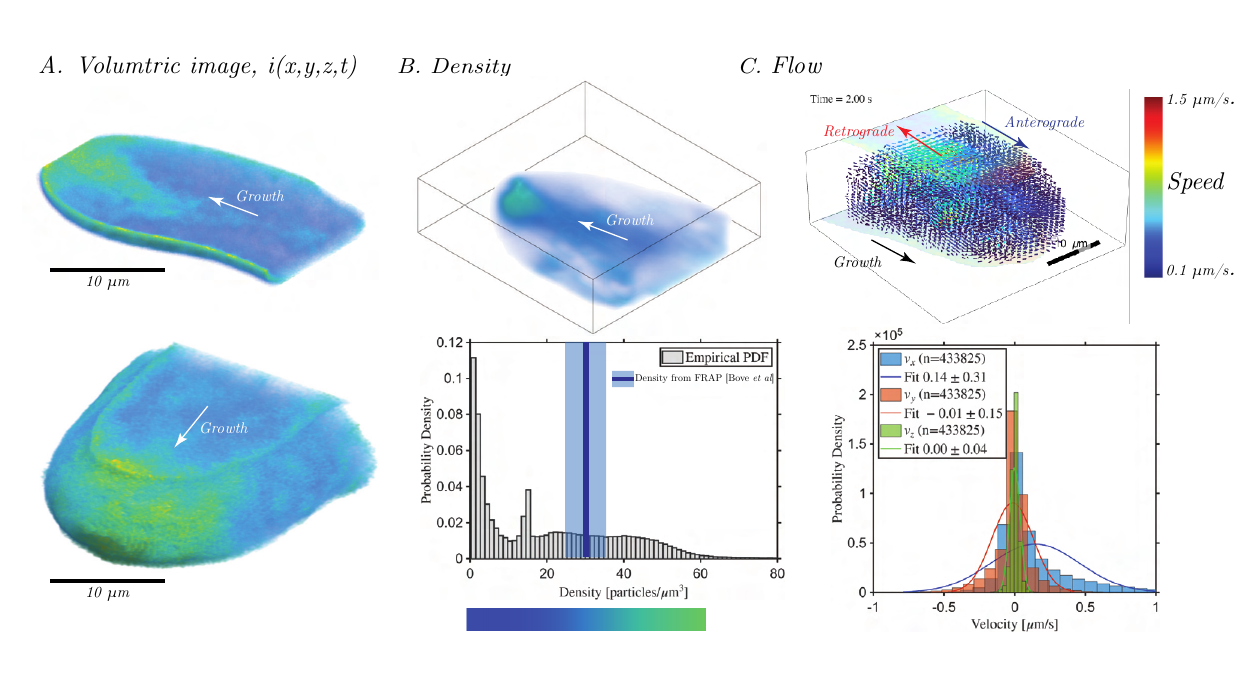}
  \caption{\textbf{Biological benchmark for vSTICS using vesicle labelling of \textit{Camellia japonica} pollen tubes.}
  \textbf{(A)} A volumetric time series of the pollen tube, displayed  in two perspectives with the sagittal plane relative to the growth axis (\textit{top}). The \textit{bottom} panel displays the same growing tube, with the membrane segmented away to only display the cellular interior (analyzed, with parameters $\text{ROI}_{xy} = 64 \ \text{pix}$, $\text{ROI}_{z} = 16 \ \text{pix}$, $\text{ROI Shift}_{xy} = 8 \ \text{pix}$, $\text{ROI Shift}_{z} = 8 \ \text{pix}$, $\text{TOI} = 5 \ \text{frames}$, $\text{TOI Shift} = 1 \ \text{frame}$.
  \textbf{(B)} Volumetric density mapping of the vesicles in the pollen tube at the first $\text{TOI}$ (\textit{top}) with the density histogram averaged over all $\text{TOIs}$ with a Gaussian fit.  
  \textbf{(C)} Velocity flow field at the first $\text{TOI}$ superimposed on a volumetric rendering of the pollen tube (\textit{top}). The corresponding normalized histogram of the velocity measured in each orthogonal direction, with a bimodal Gaussian fit for components $v_{x}$ and $v_{y}$, and a Gaussian fit for $v_{z}$.
  }
  \label{fig:vSTICSCJ-vesicle}
\end{figure}

Cargo vesicles that carry the polysaccharides needed to supply structural components to maintain continuous cell wall growth are trafficked to the growing apex in the anterograde region of the pollen tube. At the growing tip, there exists a delicate balance of cargo vesicles that are used to elongate the cell wall by fusion of the vesicle membrane to the existing cell membrane, and its contents are used to polymerize new components of the cell wall and membrane \cite{Chebli2013}, visualized by the higher fluorescence intensity at the membrane of growing pollen tube (Fig. \ref{fig:vSTICSCJ-vesicle}, \textbf{A} \textit{top}). To analyze the interior vesicles, the membrane was removed by segmentation using Ilastik \cite{Berg2019Ilastik} (Fig. \ref{fig:vSTICSCJ-vesicle}, \textbf{A} \textit{bottom}). The recycling of used and unused vesicles along the retrograde region is marked by a characteristic growth cone, displayed in Fig. \ref{fig:vSTICSCJ-vesicle}, \textbf{A} \textit{top}. Cargo vesicles are sub-diffraction-limited in size (approximately 100~nm in diameter) and present at much higher densities than the organelles, making tracking with any variation of SPT or PIV effectively impossible. Here, intensity fluctuation analysis provides a robust way of determining densities and flow. Conditions with sub-diffraction-limited particles at high density and overlapping trajectories are representative of a broad class of secretory and trafficking systems for which vSTICS is particularly advantageous. Density measurements for the vesicles are shown in Fig. \ref{fig:vSTICSCJ-vesicle} \textbf{B}, \textit{top}, with a clear higher density region located near the growing tip, and along the retrograde regions. Vesicle densities are approximately tenfold higher than mitochondrial densities
(Fig.~\ref{fig:vSTICSCJ-vesicle}\textbf{B}, \textit{bottom}), 
with values of $\sim 3$ vs.\ $30 \ \text{particles}\,\mu\text{m}^{-3}$ for mitochondria and vesicles, respectively. Flow vector fields are also recovered with anterograde (blue) and retrograde (red) regions separated by slow and fast regions respectively (Fig. \ref{fig:vSTICSCJ-vesicle} \textbf{C}, \textit{top}). As shown previously in mitochondrial mapping, a bimodal distribution is recovered in the velocity histograms, as is expected  for the bimodal transport pathways in pollen tubes (Fig. \ref{fig:vSTICSCJ-vesicle} \textbf{C}, \textit{bottom}). Density estimates are on the same order of magnitude ($\approx 30 \ \mu\text{m}^{-3}$) for the apex region as previously reported using FRAP in lily pollen tubes \cite{Bove2008VesicleDynamicsPollenTubes}.

\section{Discussion}\label{Discussion}

% ----------------------------------------------------
% 1) Methodological advances and validation
% ----------------------------------------------------
\subsection*{Methodological advances and validation}

By extending spatio-temporal image correlation spectroscopy into the volumetric domain and pairing it with rapid lattice light-sheet imaging, 
our vSTICS framework provides a general way to measure particle \emph{density}, slow \emph{diffusion}, and \emph{directed transport} at voxel scale in living specimens. An asymmetric 3D Gaussian model of the spatio-temporal ACF can be fit to the space-time correlation function to yield: (i) the zero-lag amplitude (number/density), (ii) the amplitude decay across time lags (diffusion), and (iii) the peak displacement (velocity), enabling simultaneous parameter recovery from the same data block volume (see workflow in Fig.~\ref{fig:overview}). This single-pass analysis underpins all results presented here.

We validated vSTICS using computer-simulated 3D time series with prescribed ground-truth diffusion and flow parameters and a specified signal-to-noise level. With ground-truth settings of $D = 10^{-3}\,\mu\text{m}^2\,\text{s}^{-1}$ and velocities of order $3\,\mu\text{m}\,\text{s}^{-1}$ per axis, vSTICS recovered both transport modes with high fidelity, as reflected in narrow error bands on the fitted decay curves and drift-derived velocity components (Fig.~\ref{fig:simulations}) with sufficient spatial and temporal sampling of the ROI-TOI analysis volume. Experimental calibrations further confirmed quantitative readouts: in stationary bead stacks, the vICS autocorrelation peak amplitude and width reported particle number and PSF dimensions consistent with independent measurements; in weakly viscoelastic gels, vTICS reported the expected diffusion dependent amplitude decay and yielded $D$ in agreement with imaging-FCS benchmarks (Fig.~\ref{fig:fluorospheres}, Fig. S9. Field-synthesis LLSM minimized photobleaching and phototoxicity \cite{Chang2019FieldSynthesis}, permitting rapid, extended 3D image volumes necessary for robust correlation statistics and stable parameter maps in living plant pollen tubes.

% ----------------------------------------------------
% 2) Biological insights
% ----------------------------------------------------
\subsection*{Pollen tube flow: reverse-fountain transport is asymmetric and largely transverse}

Applying vSTICS to mitochondrial dynamics in growing \emph{Camellia japonica} pollen tubes refines the classic ``reverse-fountain'' model \cite{Bove2008VesicleDynamicsPollenTubes,Rounds2011, Chebli2013, Hepler2001}. Velocity histograms and vector fields reveal a clear asymmetry: anterograde motion is slower (typically $0.1 \ \text{- 1} \ \mu\mathrm{m}\,\mathrm{s}^{-1}$) whereas retrograde transport is substantially faster, peaking near $3\,\mu\mathrm{m}\,\mathrm{s}^{-1}$ (Fig.~\ref{fig:vSTICSCJ}: \textbf{D}). The distributions are bimodal in the transverse components and narrowly centered near zero axially, indicating transport predominantly confined to transverse planes with negligible $v_{z}$ ($0\pm0.1\,\mu\mathrm{m}\,\mathrm{s}^{-1}$). Together, these observations support an \emph{asymmetric transverse} circulation with $v_{\mathrm{retro}}/v_{\mathrm{antero}}\approx 3$, updating prior symmetric flow assumptions \cite{Liu2017FountainStreaming}.

\begin{figure}[t]
  \centering
  \includegraphics[width=\textwidth]{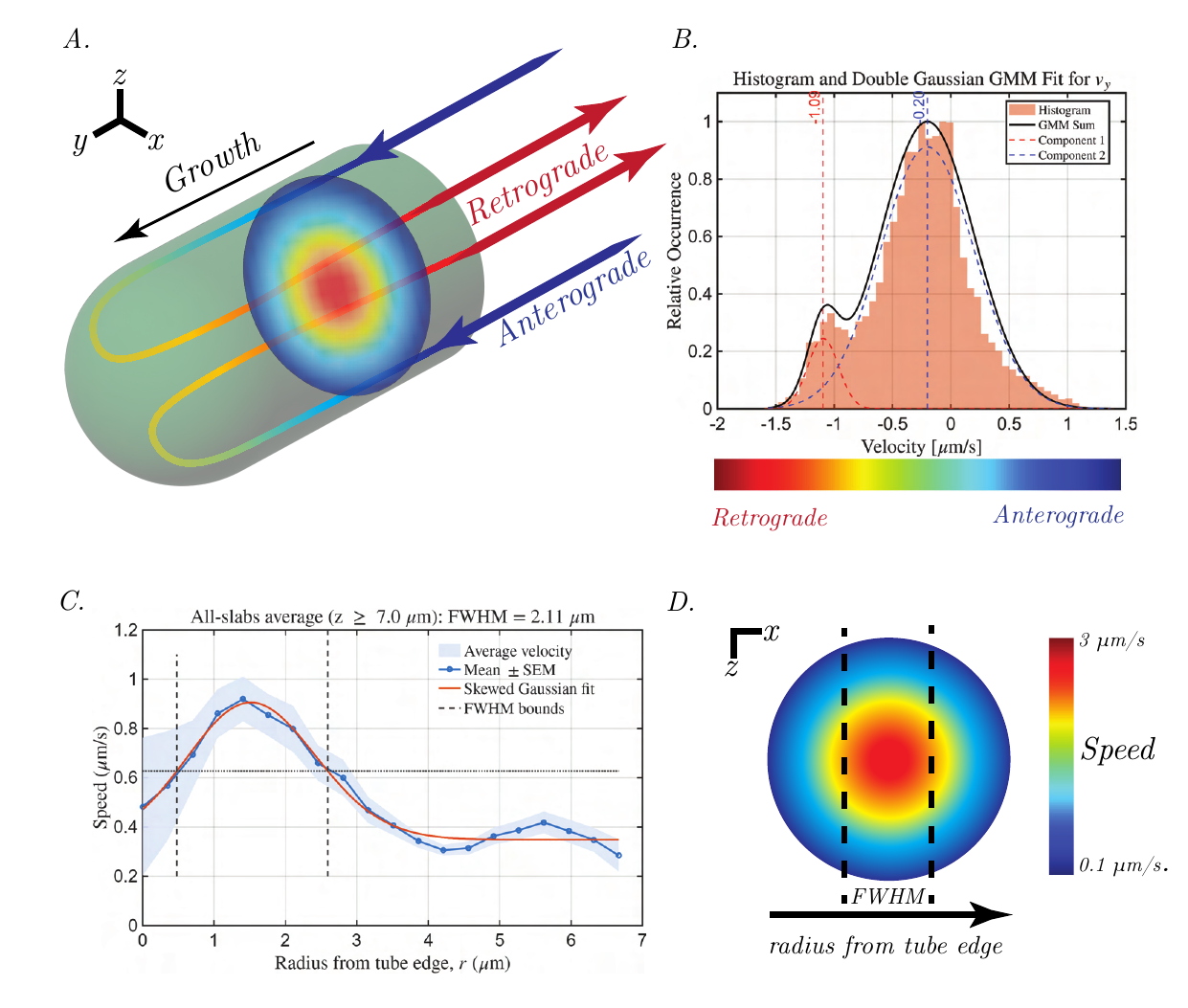}
  \caption{\textbf{Improved model for pollen tube growth mechanisms.}
  \textbf{(A)} A generalized growing pollen tube with the transverse ($xz$ plane) velocity distribution visualized at the bottom right as an asymmetric colour gradient (\textit{bottom right}) derived from the bimodal Gaussian velocity $v_{y}$ fit \textbf{(B)}. \textbf{(C)} The velocity of the mitochondria, plotted as a function of distance away from the edge of the growing pollen tube ($xz$ plane), with the blue shaded bars describing the variance across the growing axis of the tube ($y$ axis).  The vertical dotted lines display the $\text{FWHM}$ bounds, as reflected in the colour gradient (\textbf{D}), representing the anterograde region of $\sim 2 \ \mu\text{m}$.
  }
  \label{fig:TubeModel}
\end{figure}

The spatial organization of this asymmetry is quantifiable. Transverse speed profiles yield a skewed-Gaussian retrograde span with $\mathrm{FWHM} =  2.2 \pm 0.2\,\mu\mathrm{m} \ (n_{\text{samples}} =5)$; for tubes of diameter $10\pm1\,\mu\mathrm{m}$, the anterograde corridor occupies roughly $\sim 20\%$ of the diameter (Fig.~\ref{fig:TubeModel}: \textbf{C}). This is consistent with the flow-field mapping (Fig.~\ref{fig:vSTICSCJ}\textbf{D}, top) and the bimodal transverse velocity fits (Fig.~\ref{fig:TubeModel}: \textbf{B}). Our measurements motivate models that incorporate $v_{\mathrm{retro}}/v_{\mathrm{antero}}\approx 3$ with minimal axial flow, and incorporate these transverse geometric constraints when predicting tip-growth dynamics.

% ----------------------------------------------------
% 3) Density & diffusion landscapes
% ----------------------------------------------------
\subsection*{Density and diffusion landscapes imply local transport regimes}

Volumetric density maps averaged across time windows show higher particle densities centrally ($\sim 6$\,particles\,$\mu\text{m}^{-3}$) tapering toward the periphery (mean $\sim 5\pm1$\,particles\,$\mu\text{m}^{-3}$, with peripheral values $\sim 4$\,particles\,$\mu\text{m}^{-3}$). Diffusion coefficients span 0.1 -- 1\,$\mu\text{m}^2\,\text{s}^{-1}$, with larger values peripherally and smaller values toward the center (Fig.~\ref{fig:vSTICSCJ}: \textbf{B},\textbf{C}). In combination with the velocity field, these gradients are consistent with a flow-dominated regime (Péclet number $\mathrm{Pe}\gtrsim 1$) near the core and relatively more diffusive behavior toward the edges---precisely where anterograde motion is concentrated. This pattern suggests that cytoskeletal and rheological architectures impose transverse compartmentalization of advection and stochastic motion.

\subsection*{vSTICS can measure dynamics of high density, diffraction limited objects}

In biological samples where there is a high density of dynamic fluorescently labelled particles, as is the case of cargo vesicles, intensity fluctuation analysis remains robust, since localization-based methods cannot be employed. With membrane labelling of the vesicles, density quantification of the characteristic growth cone, along with the transport highways are extracted simultaneously. The decay of the correlation function amplitude was insignificant, so the diffusion coefficient is not reported.

% ----------------------------------------------------
% 4) Practical guidance
% ----------------------------------------------------

\subsection*{Practical guidance: parameterization and sampling}

The analyses in this Supplementary Information identify two practical considerations for robust vSTICS measurements. \emph{First}, ROI and TOI tiling must balance spatial resolution against correlation signal-to-noise ratio. In our datasets, $\mathrm{ROI}_{xy}=64$\,px, $\mathrm{ROI}_{z}=16$\,px, shifts of 5 and 2\,px, and $\mathrm{TOI}=5$ frames with unit shift provided a good compromise between map density and fit stability. \emph{Second}, estimating $D$ from amplitude decay, rather than relying only on time-dependent peak broadening as in iMSD \cite{iMSD}, improved robustness in strong-flow regions because decorrelation at the PSF length scale remained accessible at the LLSM volume rate.

More broadly, the results presented here show that recovered parameters depend on detector noise, background normalization, ROI size, axial sampling, temporal windowing, and rolling $z$-stack acquisition. Accordingly, simulation-based characterization should be treated as a first step when adapting vSTICS to a new imaging platform or specimen.

% ----------------------------------------------------
% 5) Limitations & improvements
% ----------------------------------------------------
\subsection*{Limitations and areas for improvement}

The PSF anisotropy remains a key limitation: large axial structure factors ($\kappa=\omega_z/\omega_{xy}$) reduce sensitivity to $v_z$, inflate uncertainty in axial peak localization, and bias velocities toward the imaging plane. Although field-synthesis LLSM mitigates anisotropy in calibration samples, in situ refractive-index mismatch and scattering can enlarge $\kappa$. Tighter light-sheet confinement, deconvolution with measured PSFs, and smaller $\mathrm{ROI}_z$ with increased axial sampling can improve axial sensitivity.

More broadly, quantitative interpretation assumes locally stationary dynamics within each TOI and adequate fluctuation sampling. When growth or flow varies rapidly, shorter TOIs and higher volumetric imaging rates are needed; conversely, for slow diffusion, longer TOIs can improve the $D$ estimates, but requires mechanical and sample stability over the sampling time. Parameter selection (block sizes, overlaps, and time-lag sampling) guided by simulation in conjunction with fluorescence imaging choices provide a practical basis for tuning vSTICS.

% ----------------------------------------------------
% NEW: Broader applications of vSTICS
% ----------------------------------------------------
\subsection*{Broader applications of vSTICS}

Beyond pollen tubes, vSTICS is well suited to 3D transport problems in slender or convoluted geometries such as axons, cilia, glandular ducts and microvascular networks, where velocities vary over tens of micrometers and particle densities preclude reliable tracking \cite{karamched2015axonalLength, vankrugten2022ocr2Cilia, su2022salivaryDuct, hill2022airwayMucus}. In these systems, volumetric imaging combined with localized 3D correlation can map flow and diffusion fields without resolving individual trajectories, enabling direct comparison to mechanochemical and hydrodynamic models that are inherently three dimensional.

More generally, the vSTICS workflow---simulation-guided parameter choice, volumetric correlation, and voxel-resolved maps of flow, diffusion and density---provides a transferable template for quantitative 3D transport mapping in any system where fluorescence can report on moving cargos or tracer particles. Tight integration between imaging design (volume rate, field of view, PSF engineering) and vSTICS analysis should enable hypothesis-driven experiments in a wide range of cellular and tissue contexts.

% ----------------------------------------------------
% 6) Conclusion
% ----------------------------------------------------

\subsection*{Conclusion}

Volumetric spatio-temporal image correlation spectroscopy (vSTICS) advances correlation methods from single plane 2D measurements to measuring transport coefficients and mapping intracellular transport and densities in 3D systems. By extracting voxel-resolved density, diffusion, and directed flow from the same volumetric microscopy time series, vSTICS provides a single-pass, quantitative description of transport that we validated with simulations and microsphere calibrations and then applied to characterize dynamics in living pollen tubes. We obtained quantitative 3D measurements consistent with an asymmetric reverse-fountain model, revealing a predominantly transverse circulation, and establishes geometric and kinematic constraints that can inform elaboration of 3D mechanochemical and hydrodynamic theories of tip growth for this inherently 3D dynamic system.

In combination with volumetric imaging modalities such as lattice light-sheet microscopy, the vSTICS workflow and accompanying parameterization guidelines (for block size, overlap, and time-lag sampling) make the method readily transferable to other crowded 3D environments where the balance of advection and diffusion governs function. While axial sensitivity can still be limited by PSF anisotropy and assumptions of local stationarity, these constraints are addressable with deconvolution, tighter light-sheet confinement, finer axial sampling, and adaptive time-windowing. Taken together, vSTICS establishes a practical and general approach for quantitative 3D transport mapping from volumetric fluorescence microscopy imaging data, enabling direct comparison to theory and guiding future model development across diverse biological systems.

\section{Materials and Methods}\label{Methods}

\subsection*{Computer simulations}

A simulation program for a volumetric intensity time series was written in Matlab 2025a \cite{MATLAB2025a} to validate vSTICS. Several experimental and instrumental parameters can be set such as particle density, beam profile (PSF), detector-based noise characteristics, diffusion coefficient, flow velocity, image size, pixel size, volume acquisition time, and length of time series. For all simulations, particles were initially seeded randomly in the volumetric pixel space at $t =0$ at the single pixel level, such that if a particle was seeded, the pixel received a value of 1, and 0 otherwise. This binary volumetric image was convolved with the Gaussian PSF, and noise was optionally added. Particle movement in time depended on the type of dynamics (flow, diffusion, or a combination) and obeyed periodic boundary conditions to mimic an infinite space.

\subsection*{Fluorosphere preparations}

Fluorescent microspheres with a diameter of 50 nm were purchased from Molecular Probes (Eugene, OR) with an absorption/emission peak of 505/515 nm. A solution was created by diluting the manufacturer concentration of  fluorospheres by a factor of 100 in MilliQ water, and vortexed prior to each use. The 100$\times$ diluted solution was then mixed with Gelrite (G1910, Sigma Aldrich) to achieve a final concentration of 0.3$\%$ Gelrite by volume to achieve a dynamic viscosity of $\eta = 8 \ \text{cP}$. The fluorosphere solution was pipetted into a custom bag (0.5 cm $\times$ 2 cm) created with a fluorinated ethylene propylene (FEP) film (50A Dupont Teflon, American Durafilm, Holliston, MA) and an impulse heat sealer. The FEP sheet was 12.7 $\mu$m thick, with a refractive index comparable to water ($n = 1.34$), minimizing optical aberrations during imaging. The bag was held in place using a custom light-sheet holder (described below) for volumetric imaging via lattice light-sheet microscopy. 

\subsection*{Pollen tube preparations for mitochondrial and vesicle labelling}

\textit{Camellia japonica} pollen was collected from the Jardin botanique de Montréal (accension 1500-71-1982). After harvesting, the pollen grains were divided into gel capsules and placed in a container with desiccant for 24 hours and  stored at -20$^{\circ}$C in falcon tubes following desiccation. Prior to imaging, pollen grains were placed on a glass coverslip in a sealed container with water-soaked tissues to hydrate for 2 hours. For \textit{in vitro} fertilization, pollen grains were deposited on a custom sample holder \cite{Rouger2016JBO_LowCostLSM} filled with a germination medium of 1.62 mM $H_{3}BO_{3}$, 2.54 mM $Ca(NO_{3})_{2}\cdot4H_{2}O$, 0.81 mM $MgSO_{4}\cdot7H_{2}O$, 1 mM $KNO_{3}$, 8 $\%$ sucrose (w/v), and 1$\%$ agar (A1296, Sigma Aldrich) in distilled water. Following deposition, a 1$\%$ low melting point agarose (A4018, Sigma Aldrich) in germination medium mixed with MitoTracker Green (Molecular Probes, Invitrogen) for mitochondrial labelling or FM 1-43 Dye (N-(3-Triethylammoniumpropyl)-4-(4-(Dibutylamino) Styryl) Pyridinium Dibromide) (Invitrogen) for vesicle labelling, to a final concentration of 2 $\mu$g/mL was gently pipetted over the pollen grains. Pollen tubes were allowed to grow for 1.5 hours and imaged via lattice light-sheet microscopy.

\subsection*{Lattice light-sheet microscopy via field synthesis}

A custom home-built lattice light-sheet microscope (LLS) via field synthesis was built for volumetric imaging. The LLS was built by combining the optical architecture of the previously published multi-modal light-sheet \cite{Rouger2016JBO_LowCostLSM} and the implementation of the field synthesis theorem \cite{Chang2019FieldSynthesis}. The following provides a brief description of the optical pathway. A laser box with five OBIS diode lasers (405, 445, 488, 561, 604 nm, Coherent Inc., Santa Clara, California) were combined using reflective and dichroic mirrors, and the laser light was focused onto a fiber output (OZ Optics Ltd., Ottawa, ON, Canada) and directed towards the excitation arm. After passing through a 5$\times$ beam expander, and circular iris, the beam is shaped to a line via a cylindrical lens (ACY254-050-A, Thorlabs). A galvanometer (galvo mirror, 8315K, Cambridge Technology) placed conjugate to the sample plane is used with an achromatic lens (AC254-075-A, Thorlabs) to scan the line over a custom annular mask with an $NA_{inner} = 0.438$ and $NA_{outer} = 0.539$. The pattern generated is similar to a square lattice, with the central line removed (Fig. S10). A telecentric relay of two tube lenses (ITL-200, Thorlabs) is used to project the illumination pattern onto the back focal plane of the illumination objective CFO Apo 40× (0.8 NA, Nikon). 

The detection path is orthogonal to the illumination path and utilizes the body of a Nikon FN-1 upright microscope (Nikon Canada Inc., Mississauga, ON, Canada). Emission photons are collected by a CFO Apo 40× (0.8 NA, Nikon), with a quad-band filter fluorescent cube (Chroma Technology Corp., Bellows Falls,
Vermont), and detected on a 1200 $\times$ 1200 Andor SONA sCMOS camera (SnowHouse Solutions Inc., Lac-Beauport,
Quebec). Excitation, sample movement, and detection, was controlled using custom LabVIEW code and a NIDAQ acquisition card (NI-PCIe-6353, National Instruments, Toronto, ON, Canada).

\subsection*{Volume autocorrelation analysis}

Volume time series were converted from the LabView binary output and viewed with ImageJ \cite{Schindelin2012Fiji}. All code, including volumetric autocorrelation, nonlinear least squares Gaussian function fitting, parameter calculations (diffusion, flow, and density), and graphical output, were implemented via Matlab 2025a \cite{MATLAB2025a} on a 4.0 GHz Dell Optiplex Tower.

\section{Theory}\label{Theory}

\subsection{Mathematical formalism of vSTICS}

Fluorescence fluctuation correlation analysis methods have been applied extensively in 1D such as FCS \cite{Magde1974FCS2, AbuArish2010BicoidFCS, Schwille1997FCCS} and in 2D with imFCS \cite{Bag2014IFFS} and the ICS suite including TICS, STICS, and kICS \cite{Petersen1993ICS, Kolin2006TICS, Hebert2005STICS, Kolin2006kICS}. Each technique involves the correlation in space or reciprocal space and/or time of fluorescence intensity fluctuations or of fluorescence photon counts collected using a fluorescence microscope. To outline the theory of 3D volumetric spatio-temporal image correlation spectroscopy (vSTICS), we start by defining the generalized time dependent normalized spatial intensity fluctuation correlation function based on the molecular detection efficiency. This base formalism assumes a single species, a single excitation laser line, and a single detection channel, but can be readily extended to cross-correlation analysis with multicolour fluorescence imaging.

\subsubsection{Normalized spatial intensity fluctuation correlation function}

The molecular detection efficiency is defined as, $\text{MDE}(\vec{r}) = \text{PSF}_{ill}(\vec{r})\text{PSF}_{det}(\vec{r})$, where $\text{PSF}_{ill}(\vec{r})$ and $\text{PSF}_{det}(\vec{r})$ are the spatially dependent point spread functions as a result of the illumination and detection schemes, respectively. This is incorporated into the standard definition of the normalized autocorrelation of the intensity fluctuations, $\delta i(x,y,z) = i(x,y,z) - \langle i \rangle_{x,y,z}$, 

\begin{equation} \label{autocorr3D}
    g(\xi, \eta, \phi) = \frac{\langle \delta i(x,y,z)\cdot \delta i(x+\xi,y+ \eta,z+\phi) \rangle}{\langle i(\vec{r}) \rangle ^2}
\end{equation}

\noindent where the angular braces denote spatial correlation (numerator) spatial averaging (denominator), and $(\xi, \eta, \phi)$ are spatial lags (pixel shifts) for the orthogonal Cartesian image coordinates. The $\text{MDE}$ transforms a spatial concentration distribution of fluorescent molecules into a corresponding spatially dependent fluorescence signal (which is described for the temporal case in \cite{Wachsmuth2003}). The normalized spatial autocorrelation function is obtained, 

\begin{equation} \label{spaceACF}
    g(\xi, \eta, \phi) = \frac{\langle \int \int \text{MDE}(x,y,z) \cdot \text{MDE}(x + \xi,y + \eta,z + \phi) \enspace dV dV' \rangle}{\langle c(\vec{r}) \rangle \langle \int \text{MDE}(\vec{r}) \enspace dV\rangle} = \frac{G(\xi, \eta, \phi)}{\langle c(\vec{r}) \rangle \langle \int \text{MDE}(\vec{r}) \enspace dV\rangle}
\end{equation}

where $G(\xi, \eta, \phi)$ is the non-normalized spatial intensity autocorrelation function, which is defined as the autocorrelation of the $\text{MDE}(\vec{r})$ function. This general form allows for derivation of the correlation function for any illumination and detection scheme. The non-normalized intensity autocorrelation is calculated via the Wiener–Khinchin theorem as the inverse Fourier transform of the power spectrum of the image: 

\begin{equation}
    G(\xi, \eta,\phi) = \mathscr{F}^{-1} \left( \left[ \mathscr{F}(i(x,y,z))\right] \cdot \left[ \mathscr{F}^{*}(i(x,y,z))\right] \right)
\end{equation}

\noindent where $\mathscr{F}$ denotes the spatial Fourier transform, $\mathscr{F}^{*}$ its complex conjugate, and $\mathscr{F}^{-1}$ denotes the inverse spatial Fourier transform.

\subsubsection{Time dependent spatial autocorrelation function}

The functional form of the time dependent normalized spatial intensity autocorrelation function $g(\xi, \eta, \phi, \tau)$, where $\tau$ denotes the temporal lag (frame shifts), is retrieved by convolving Eq. \ref{spaceACF} with a time dependent probability distribution $P_{\vec{r_{\tau}}|\vec{r_{0}}}(-\vec{r}|\vec{r_{0}}(0) = 0)$, such that, 

\begin{equation} \label{generalACF}
    g(\xi, \eta, \phi, \tau) = \frac{G(\xi, \eta, \phi)}{\langle c(\vec{r}) \rangle \langle \int \text{MDE}(\vec{r}) \enspace dV\rangle} \circledast P_{\vec{r_{\tau}}|\vec{r_{0}}}(-\vec{r}|\vec{r_{0}}(0) = 0)
\end{equation}

The probability distribution $P_{\vec{r_{\tau}}|\vec{r_{0}}}$ describes the likelihood of finding a particle at $\vec{r}$ given prior knowledge of its position at some $\vec{r}_0$. The unique functional forms of $P$ are the Green’s function of the partial differential equation that describes the concentration, $c(\vec{r},t)$, and is dependent on the type of molecular transport, such as diffusion, flow, or a combination, including the possibility of multiple dynamic species (Table S1) .

\subsection{Density, diffusion, and flow}

At zero spatial and temporal lags, $(\xi, \eta, \phi, \tau) = \vec{0}$, 

\begin{equation}
    g(\vec{0}) = \frac{1}{\langle c \rangle} \frac{\langle \int \text{MDE}(x,y,z)^{2} \ \text{d}V \rangle }{ \langle \int \text{MDE}(x,y,z)^{2} \ \text{d}V \rangle^{2} } \label{Veff}
\end{equation}

The second term yields an expression for the effective observation volume (i.e. focal volume), 

\begin{equation} \label{VeffEq}
    V_{eff} = \frac{\langle \int \text{MDE}(x,y,z) \ \text{d}V \rangle^{2} }{\langle \int \text{MDE}(x,y,z)^{2} \ \text{d}V \rangle  } = \frac{\langle N \rangle}{\langle c \rangle}
\end{equation}

\noindent  which yields the expected dependence of the zero lags amplitude of the normalized autocorrelation function on the inverse of the mean number of independent fluorescent molecules in the observation volume, 

 \begin{equation}
    g(\vec{0}) = \frac{1}{\langle c \rangle}\cdot \frac{\langle c \rangle}{\langle N \rangle} = \frac{1}{\langle N \rangle}
\end{equation}

We have presented the $\text{MDE}$ function as a general function of the optical system, but in most cases, it is modeled as a 3D Gaussian to approximate the central lobe of the Airy disk function,

\begin{equation} \label{MDE}
    \text{MDE}(\vec{r}) = I_{0} \cdot \text{exp} \left( -2 \cdot \frac{x^{2}+y^{2}}{\omega_{xy}^{2}}\right)\cdot \text{exp}\left(- 2\frac{z^2}{\omega_z^2}\right)
\end{equation}

By substituting Eq. \ref{MDE} into Eq. \ref{autocorr3D}, we obtain the normalized intensity fluctuation autocorrelation function,

\begin{equation} \label{fitfunction}
    g(\xi, \eta, \phi) = g(\vec{0}) \cdot \text{exp}\left[ - \frac{(\xi - \mu_{\xi})^{2} + (\eta-\mu_{\eta})^{2}}{\omega_{xy}^{2}} - \frac{(\phi-\mu_{\phi})^{2}}{\omega_{z}^{2}}\right] + g_{\infty}
\end{equation}

\noindent where fit parameters $\mu_{\xi,\eta,\phi}$ correspond to the peak shifts in lag space for flow (Fig. \ref{fig:overview}: \textbf{E}), $g(\vec{0)}$ relates to the density and diffusion related amplitude decorrelation in time (Fig. \ref{fig:overview}: \textbf{F}), $\omega_{xy} \ \text{and} \ \omega_{z} $ are the $e^{-2}$ radii of the correlation function with meaningful interpretations at $\tau = 0$ (Fig. \ref{fig:overview}: \textbf{C}), and $g_{\infty}$ represents long range correlations that may offset the correlation function due to incomplete decay or actual long range feature correlations.

Solving Eq. \ref{VeffEq} for the effective volume yields $V_{eff} = \pi^{3/2}\omega_{xy}^{2}\omega_z$, where $\omega_{xy}$ and $\omega_{z}$ are the $e^{-2}$ radii of the measured $\text{MDE}$.

\bmhead{Acknowledgments}

P.W.W. acknowledges support from a Natural Sciences and Engineering Research Council of Canada Discovery Grant (NSERC RGPIN-2023-03975) and the Canada Foundation for Innovation. The authors are grateful to Dr. Anja Geitmann and Dr. Youssef Chebli for helpful guidance and insightful discussions on pollen tube biology.

\bmhead{Code Availability}

The code used in this study is publicly available at \url{https://github.com/mamahmood81/vSTICS}. Detailed documentation is provided in the Supplementary Information through the \textit{User Guide for the vSTICS Analysis Pipeline Script Version} and the \textit{User Guide for the vSTICS Simulation and Analysis GUI}, which describe the script-based workflow and GUI-based workflow, respectively.

\section*{Declarations}

 None.

\bibliography{sn-bibliography}% common bib file

\end{document}